\documentclass[12pt]{article}
\pdfoutput=1 % if your are submitting a pdflatex (i.e. if you have images in pdf, png or jpg format)

\usepackage{jheppub} % for details on the use of the package, please see the JHEP-author-manual
\usepackage{graphicx}  % needed for figures
\usepackage{dcolumn}   % needed for some tablese$\cdot$cm
\usepackage{bm}        % for math
\usepackage{amssymb}   % for math
\usepackage{subfigure}
\usepackage{epstopdf}
\usepackage{color}
\usepackage{slashed}
\usepackage[utf8]{inputenc}
\usepackage{multirow}
\usepackage{footnote}
\usepackage{amsmath}
\allowdisplaybreaks[4]
\usepackage{accents}
\newlength{\dhatheight}

\usepackage{hyperref}

\title{Implication of Higgs Precision Measurement on New Physics}
\author[1]{Ning Chen,}
\author[2,3]{Tao Han,}
\author[4]{Shufang Su,}
\author[4,5,6]{Wei Su,}
\author[7]{Yongcheng Wu$^{*}$\note[*]{Speaker}}

\affiliation[1]{School of Physics, Nankai University, Tianjin 300071, China}
\affiliation[2]{Department of Physics and Astronomy, University of Pittsburgh,  Pittsburgh, PA 15260, USA}
\affiliation[3]{Department of Physics, Tsinghua University, and Collaborative Innovation Center of Quantum Matter, Beijing, 100086, China}
\affiliation[4]{Department of Physics, University of Arizona, Tucson, Arizona  85721, USA}
\affiliation[5]{CAS Key Laboratory of Theoretical Physics, Institute of Theoretical Physics, Chinese Academy of Sciences, Beijing 100190, China}
\affiliation[6]{School of Physics, University of Chinese Academy of Sciences, Beijing 100049, China}
\affiliation[7]{Ottawa-Carleton Institute for Physics, Carleton University, 1125 Colonel By Drive, Ottawa, Ontario K1S 5B6, Canada}

\emailAdd{ustc0204.chenning@gmail.com, than@pitt.edu, shufang@email.arizona.edu, weisv@itp.ac.cn, ycwu@physics.carleton.ca}

\abstract{
Future precision measurements of the Standard Model (SM) parameters at the proposed Z-factories and Higgs factories may have significant impacts on new physics beyond the Standard Model (BSM). We illustrate this by focusing on the Type-II two Higgs doublet model. A multi-variable global fitting is performed with full one loop contributions to relevant couplings. The Higgs signal strength measurement at proposed Higgs factories can provide strong constraints on new physics and are found to be complementary to the Z-pole measurements. 
}

\begin{document}
\titlepage
\maketitle
\newpage

\flushbottom

\section{Introduction}
All the indications from the current measurements seem to confirm the validity of the Standard Model (SM) and the observed Higgs boson is SM-like. However, there are compelling arguments, both from theoretical and observational points of view, in favor of the existence of new physics beyond the Standard Model (BSM)~\cite{Giudice:2008bi}. One of the most straightforward, but well-motivated extensions is the two Higgs doublet model (2HDM)~\cite{Branco:2011iw} which has five massive scalars ($h$, $H$, $A$, $H^\pm$) after electroweak symmetry break (EWSB). 

Complementary to the direct searches which has been actively carried out at the LHC, precision measurements of the Higgs properties could also lead to relevant insights towards new physics. The proposed Higgs factories (CEPC~\cite{CEPC-SPPCStudyGroup:2015csa, CEPCStudyGroup:2018ghi}, ILC~\cite{Baer:2013cma} or FCC-ee~\cite{Gomez-Ceballos:2013zzn,fccpara,fccplan}) can push the measurement of Higgs properties into sub-percentage era. Using the Type-II 2HDM as example, we perform a global fit to get the prospects of these measurements in constraining the model parameter spaces~\cite{Chen:2018shg}.

\section{Higgs and Electroweak Precision Measurement}
\subsection{Higgs Measurements}
\begin{figure}[!hbt]
\centering
\includegraphics[width=0.47\textwidth]{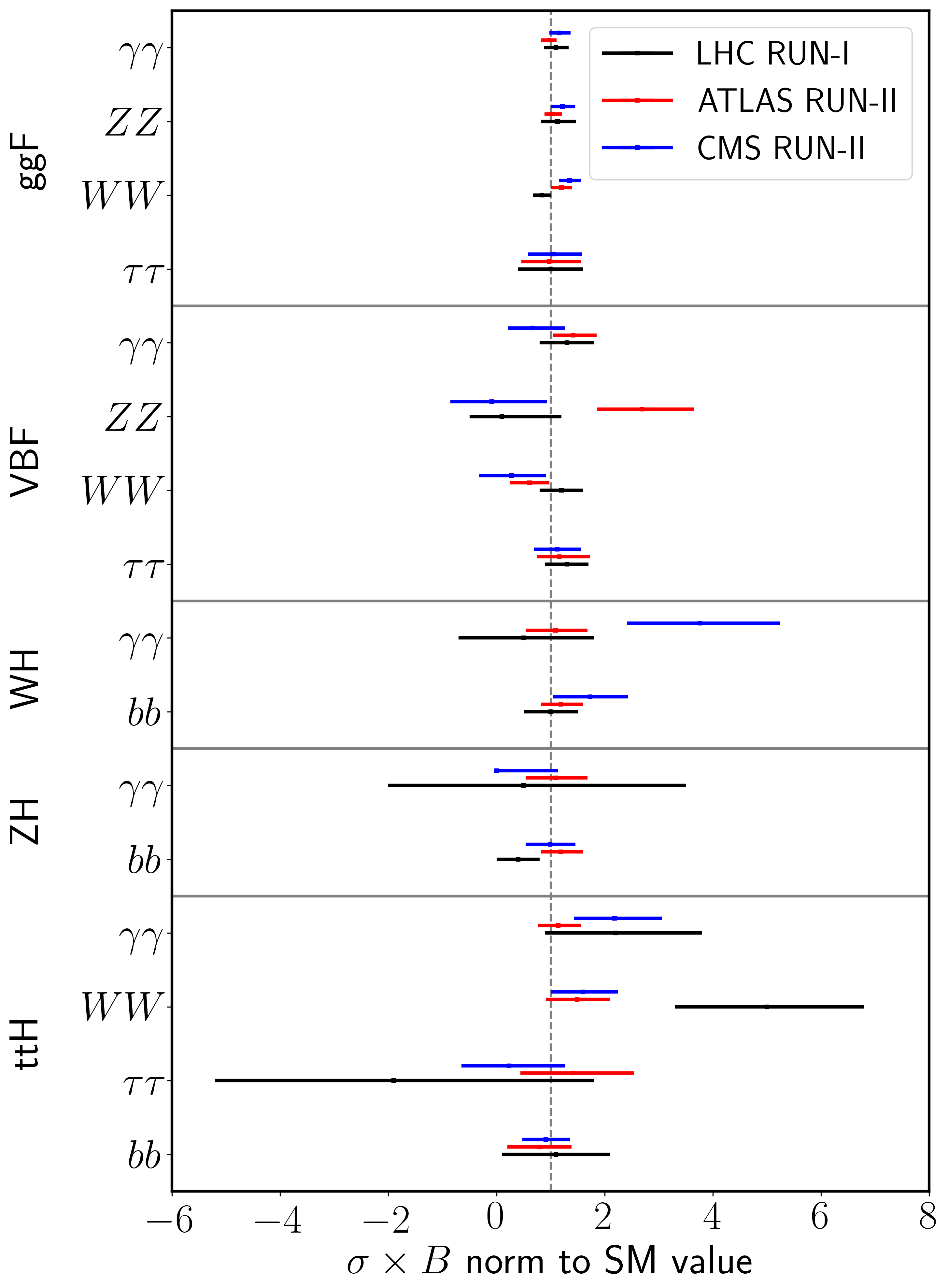}
\includegraphics[width=0.43\textwidth]{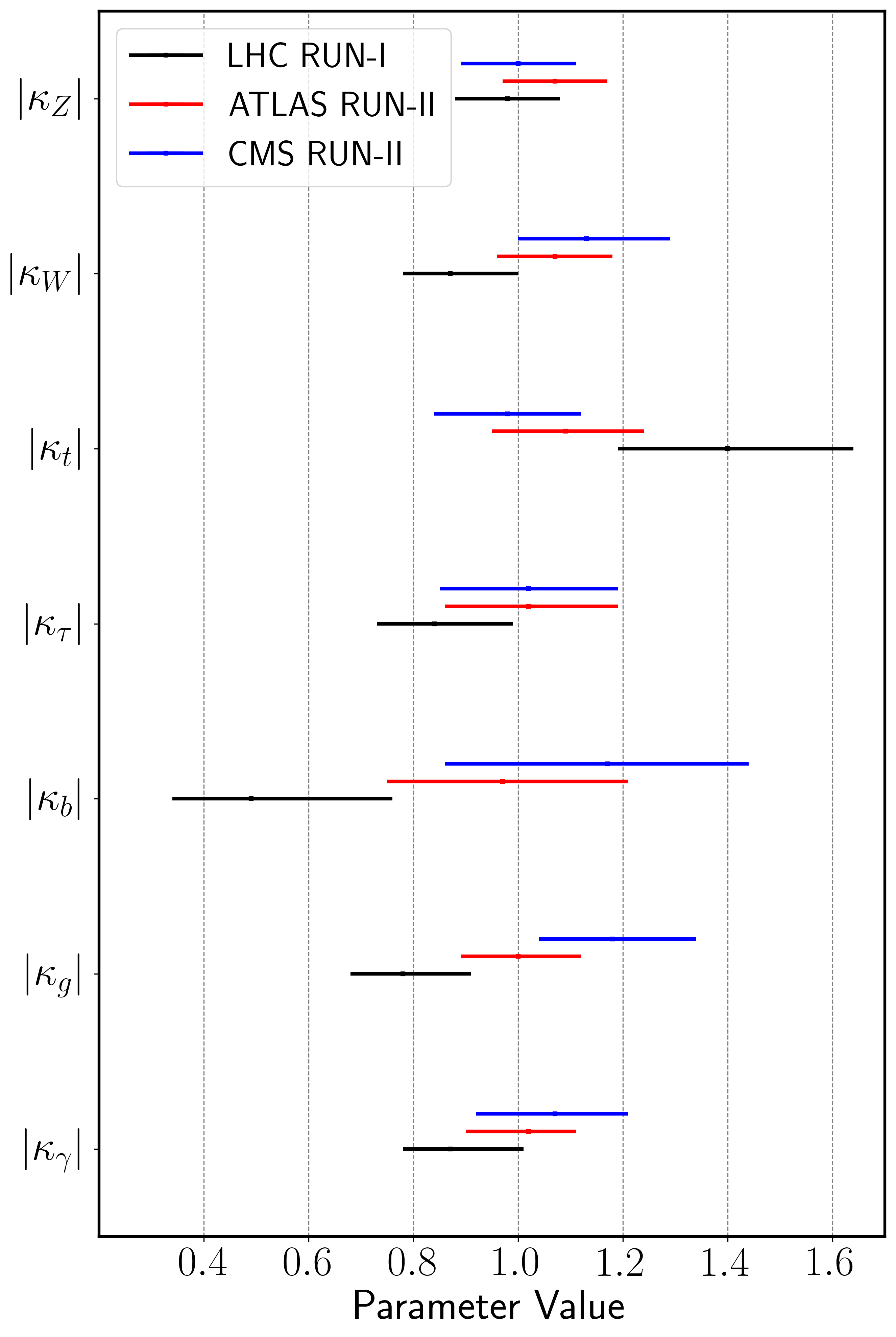}
\caption{Current Higgs signal strength measurements for various channels (left) and the constraints on the $\kappa$ of different couplings assuming no additional BSM channel for the Higgs (right). Data is retrieved from~\cite{Khachatryan:2016vau,ATLAS-CONF-2018-031,Sirunyan:2018koj}. }
\label{fig:LHCHiggsSS}
\end{figure}

Currently, the ATLAS and CMS have carried out the comprehensive measurements of the Higgs signal strength in various channels with 7 and 8 TeV combined data~\cite{Khachatryan:2016vau} as well as the 13 TeV data~\cite{Sirunyan:2018koj,ATLAS-CONF-2018-031}. The results are collected in Fig.~\ref{fig:LHCHiggsSS}. In the left panel, we show the signal strength defined as
\begin{align}
\mu_i = \frac{\left(\sigma\times{\rm Br}\right)_i}{\left(\sigma\times{\rm Br}\right)_{\text{SM}}}
\end{align}
for different production and decay channels. While the right panel shows the constraints from the signal strength measurement on the $\kappa_i$ which is defined as 
\begin{align}
\kappa_i = \frac{g_{hii}}{g_{hii}^{\text{SM}}}
\end{align}
In each figure, the black line represents the RUN-I combined measurement~\cite{Khachatryan:2016vau}, the red and blue lines represent the RUN-II measurements from ATLAS~\cite{ATLAS-CONF-2018-031} and CMS~\cite{Sirunyan:2018koj} respectively. From these plots, we find that in some channels (such as those in $ZH$ production mode) the precision has been improved significantly. For the coupling measurement, it still has at least 10\% uncertainties as shown in the right panel of Fig.~\ref{fig:LHCHiggsSS}.

\begin{figure}[!hbt]
\centering
\includegraphics[width=0.7\textwidth]{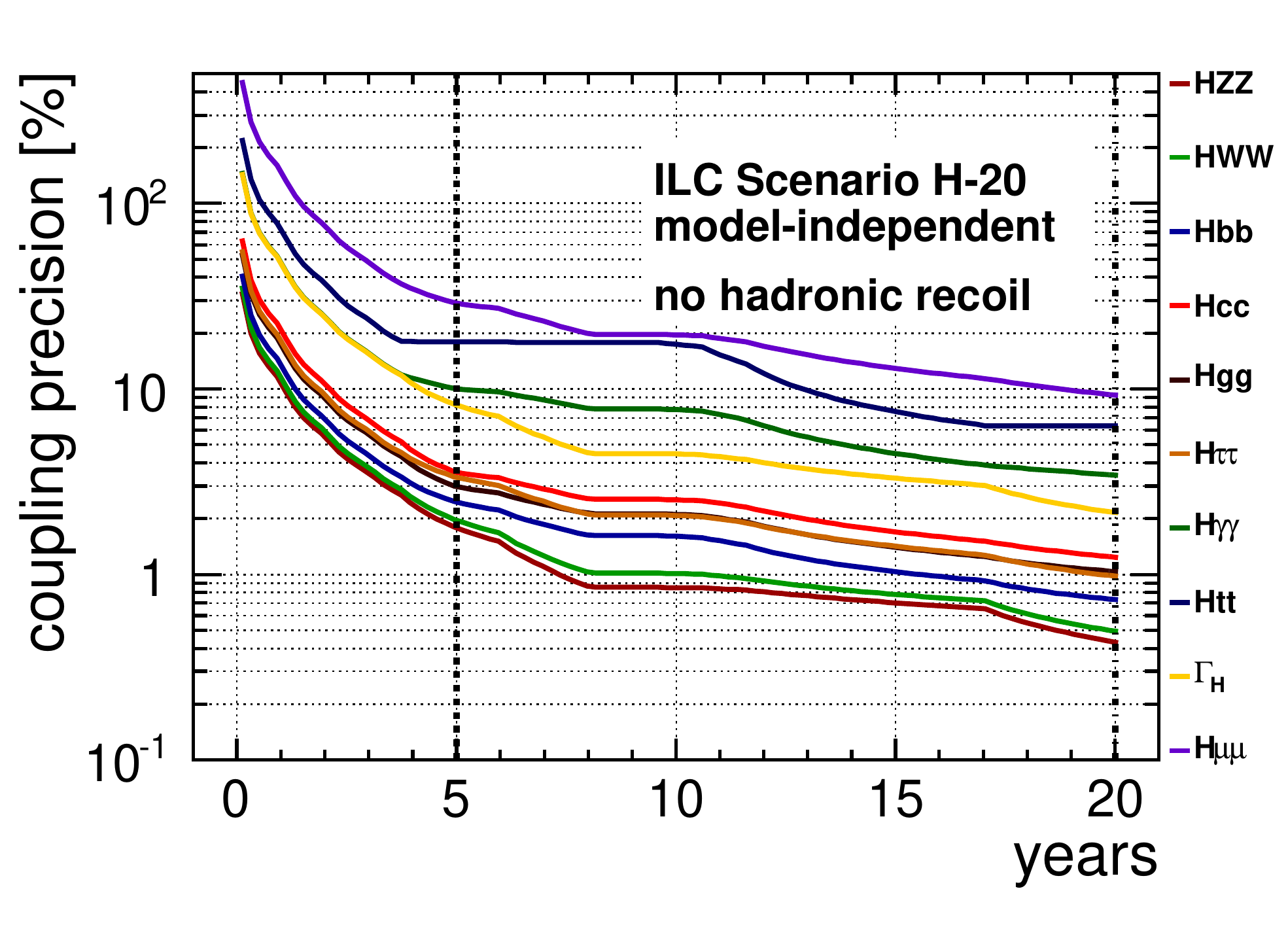}
\caption{The precision on various couplings of the Higgs boson in scenario H-20. Figure is getting from~\cite{Barklow:2015tja}.}
\label{fig:KappaH20}
\end{figure}

The measurements can be significantly improved with the proposed Higgs factories. As an example, the H-20 scenario of ILC~\cite{Barklow:2015tja} will accumulate about 4000 fb$^{-1}$ at $\sqrt{s}=$500 GeV, 200 fb$^{-1}$ at $\sqrt{s}=$350 GeV and 2000 fb$^{-1}$ at $\sqrt{s}=$250 GeV. The detailed precisions of the signal strength measurements for different channels and $\sqrt{s}$ are listed in Tab.~\ref{tab:mu_precision}. Correspondingly, the coupling measurements can be improved considerably as shown in Fig.~\ref{fig:KappaH20}. With the huge data accumulated in the Higgs factories, sub-percent precision can be expected for the measurement of the Higgs couplings which will in turn put stringent constraints on new physics.

\begin{table}[tb]
 \begin{center}
  \begin{tabular}{|l|r|r|r|r|r|r|}
   \hline
   collider&\multicolumn{6}{c|}{ILC} \\
   \hline
   $\sqrt{s}$  &  \text{250\,GeV}  &
   \multicolumn{2}{c|}{\text{350\,GeV}}  & \multicolumn{3}{c|}{\text{500\,GeV}} \\
   %\hline
   $\int{\mathcal{L}}dt $  &  $\text{2 ab}^{-1} $  &
   \multicolumn{2}{c|}{$\text{200 fb}^{-1}$}  & \multicolumn{3}{c|}{$\text{4 ab}^{-1}$} \\
   \hline
    \hline
production& $Zh$      & $Zh$     & $\nu\bar{\nu}h$     & $Zh$     & $\nu\bar{\nu}h$ & $t\bar{t}h$ \\
   \hline
  $\Delta \sigma / \sigma$ & 0.71\% & 2.1\% & $-$ & 1.06 & $-$ & $-$ \\ \hline \hline
   decay & \multicolumn{6}{c|}{$\Delta (\sigma \cdot BR) / (\sigma \cdot BR)$}  \\
  \hline
   $h \to b\bar{b}$                    &  0.42\%    & 1.67\%         & 1.67\%                  & 0.64\%    & 0.25\%           & 9.9\%        \\

   $h \to c\bar{c}$              & 2.9\%          & 12.7\%    & 16.7\%                   & 4.5\%     & 2.2\%           & $-$           \\

   $h \to gg$                       & 2.5\%           & 9.4\%    & 11.0\%                  & 3.9\%     & 1.5\%           & $-$           \\

   $h \to WW^*$      & 1.1\%          & 8.7\%    & 6.4\%                  & 3.3\%    & 0.85\%           & $-$           \\

   $h \to \tau^+\tau^-$      &2.3\%           &4.5\%          & 24.4\%                  & 1.9\%    & 3.2\%           & $-$           \\

   $h \to ZZ^*$       & 6.7\%        & 28.3\%     & 21.8\%                   & 8.8\%     & 2.9\%           & $-$           \\

   $h \to \gamma\gamma$    & 12.0\%     & 43.7\%     &50.1\%                   & 12.0\%   &6.7\% & $-$ \\

   $h \to \mu^+\mu^-$        & 25.5\%        & 97.6\%     & 179.8\%                  & 31.1\%     & 25.5\%            & $-$           \\
   \hline
  \end{tabular}
  \caption{Estimated statistical precisions for Higgs measurements obtained at the ILC with various center-of-mass energies~\cite{Barklow:2015tja}.
   }
\label{tab:mu_precision}
  \end{center}
\end{table}

\subsection{Electroweak Measurement}
Beside Higgs related measurements, the proposed Higgs factories will also have their own Z-pole measurement. The current best precision measurements for Z-pole physics mostly come from the LEP-I as well as Tevatron and the LHC~\cite{Baak:2014ora,Haller:2018nnx}. These measurements will be significantly improved at future lepton colliders~\cite{CEPC-SPPCStudyGroup:2015csa,Gomez-Ceballos:2013zzn,fccpara,fccplan,Asner:2013psa}. The anticipated precisions on several measurements for Z-pole program are listed in Tab.~\ref{tab:stu-ee-input}.

Given the complexity of a full Z-pole precision global fit, we instead use the Peskin-Takeuchi oblique parameters S, T and U~\cite{Peskin:1991sw} to present the implications of the Z-pole precision. Corresponding to the precision listed in Tab.~\ref{tab:stu-ee-input}, the constrained S, T and U ranges and associated error correlation matrix are listed in Tab.~\ref{tab:STU}, which are obtained by using {\tt Gfitter} package~\cite{Baak:2014ora}.

\begin{table}[tbh]
\centering
\setlength{\tabcolsep}{.3em}
\begin{tabular}{|c|c|}
\hline
& ILC Precision  \\
\hline
$\alpha_s(M_Z^2)$ &
$\pm 1.0 \times 10^{-4}$% ILC
\\
$\Delta\alpha_{\rm had}^{(5)}(M_Z^2) $ &
$\pm 4.7 \times 10^{-5}$ % ILC
\\
$m_Z$ [GeV] &
$\pm0.0021$ % ILC
\\
$m_t$ [GeV] (pole)&
$\pm 0.03_{\rm exp} \pm 0.1_{\rm th}$  %ILC
\\
$m_h$ [GeV] &
$<\pm 0.1$ % ILC
\\
%\hline
$m_W$ [GeV] &
$\left(\pm 5_{\rm exp} \pm 1_{\rm th}\right) \times 10^{-3}$ % ILC ~\cite{Baak:2014ora,Freitas:2013xga}~\cite{Baak:2014ora,Freitas:2013xga}
   \\
$\sin^2\theta^{\ell}_{\rm eff}$  &
$\left(\pm 1.3_{\rm exp} \pm 1.5_{\rm th}\right) \times 10^{-5} $  % ILC
   \\
$\Gamma_{Z}$ [GeV] &
%~\cite{LiangTalk,Freitas:2014hra}
$\pm 0.001$   % ILC ~\cite{AguilarSaavedra:2001rg}
%~\cite{Gomez-Ceballos:2013zzn,Freitas:2014hra}
   \\
\hline
\end{tabular}
\caption{Anticipated precisions of the EW observables at the future lepton colliders. The results are mainly from~\cite{Fan:2014vta,Lepage:2014fla,Baak:2013fwa,Baak:2014ora,LiangTalk}.
}
\label{tab:stu-ee-input}
\end{table} 

\begin{table}[tb]
\centering
%\resizebox{\textwidth}{!}{
  \begin{tabular}{|l|c|r|r|r|c|r|r|r|}
   \hline
    & \multicolumn{4}{c|}{Current ($1.7 \times 10^{7}\ Z$'s)}&\multicolumn{4}{c|}{ILC ($10^{9}Z$'s)} \\
   \hline
   \multirow{2}{*}{}
   &\multirow{2}{*}{$\sigma$} &\multicolumn{3}{c|}{correlation}
   &{$\sigma$} &\multicolumn{3}{c|}{correlation} \\
   \cline{3-5}\cline{7-9}
   & &$S$&$T$&$U$&($10^{-2}$)&$S$&$T$&$U$\\
   \hline
   $S$& $0.04 \pm 0.11$& 1 & 0.92 & $-0.68$ &  $3.53$    &   1    &    0.988   & $-0.879$ \\
\hline
   $T$&$0.09\pm 0.14$& $-$ & 1 & $-0.87$ &  $4.89$   &   $-$    &   1    &   $-0.909$\\
\hline
   $U$& $-0.02 \pm 0.11$& $-$ & $-$ & 1 &  $3.76$     &   $-$    &   $-$    & 1 \\
   \hline
  \end{tabular}
%  }
  \caption{Estimated $S$, $T$, and $U$ ranges and correlation matrices $\rho_{ij}$  from $Z$-pole precision measurements  of the current results, mostly  from LEP-I~\cite{ALEPH:2005ab},  and at future lepton colliders ILC ~\cite{Asner:2013psa}. {\tt Gfitter} package~\cite{Baak:2014ora} is used in obtaining those constraints.  }
\label{tab:STU}
\end{table}

\section{Type-II 2HDM}
\subsection{Model Setup}
In 2HDM, two $SU(2)_L$ scalar doublets $\Phi_i$ ($i=1,2$) with a hyper-charge assignment $Y=+1/2$:
\begin{align}
\Phi_i = \left(\begin{array}{c}
\phi_i^+ \\
\frac{v_i + \phi_i^0 + i G_i}{\sqrt{2}}
\end{array}\right)
\end{align}
The neutral component of each doublet obtains a vacuum expectation value (vev) $v_i$ (i=1,2) after EWSB satisfying $v_1^2+v_2^2=v^2=(246\text{ GeV})^2$, and $v_2/v_1=\tan\beta$.

The 2HDM Lagrangian for the Higgs sector can be written as
\begin{align}
\mathcal{L} = \sum_i|D_\mu\Phi_i|^2 - V(\Phi_1,\Phi_2) + \mathcal{L}_{\text{Yuk}},
\end{align}
where the Higgs potential has following form:
\begin{align}
V(\Phi_1,\Phi_2) = & m_{11}^2 \Phi_1^\dagger\Phi_1 + m_{22}^2\Phi_2^\dagger\Phi_2 - m_{12}^2(\Phi_1^\dagger\Phi_2+h.c.) + \frac{\lambda_1}{2}(\Phi_1^\dagger\Phi_1)^2 + \frac{\lambda_2}{2}(\Phi_2^\dagger\Phi_2)^2 \nonumber\\
& \lambda_3(\Phi_1^\dagger\Phi_1)(\Phi_2^\dagger\Phi_2) + \lambda_4 (\Phi_1^\dagger\Phi_2)(\Phi_2^2\Phi_1)+\frac{1}{2}\lambda_5\left[(\Phi_1^\dagger\Phi_2)^2+h.c. \right]
\end{align}
where we assume the conservation of CP symmetry. 

In general, there are four types Yukawa couplings assigning different doublet to different type fermions. In this work, we focus on Type-II of which the Yukawa couplings is:
\begin{align}
-\mathcal{L}_{\text{Yuk}}=Y_d\bar{Q}_L\Phi_1d_R + Y_e\bar{L}_L\Phi_1e_R + Y_u \bar{Q}_L i\sigma_2\Phi_2^*u_R + h.c. 
\end{align} 

After EWSB, three of the degree of freedom in two doublets are eaten by the SM gauge bosons, providing their masses. The remaining physical mass eigenstates are two CP-even neutral Higgs bosons $h$ and $H$, one  CP-odd neutral Higgs boson $A$, and a pair of charged scalars $H^\pm$. Instead of the eight parameters in the potential $m_{11}^2$, $m_{22}^2$, $m_{12}^2$, $\lambda_{1,2,3,4,5}$, more physical parameters will be used: ($v$, $\tan\beta$, $\alpha$, $m_h$, $m_H$, $m_A$, $m_{H^\pm}$, $m_{12}^2$).

In term of these mass eigenstates, the effective Lagrangian for the couplings between these scalars and SM particles is
\begin{align}
\mathcal{L}&=\kappa_Z\frac{m_Z^2}{v}Z_{\mu}Z^\mu h + \kappa_W\frac{2m_W^2}{v}W_\mu^+W^{\mu-}h+\kappa_g \frac{\alpha_s}{12\pi v}G_{\mu\nu}^aG^{a\mu\nu}h + \kappa_\gamma\frac{\alpha}{2\pi v}A_{\mu\nu}A^{\mu\nu}h \nonumber\\
&+\kappa_{Z\gamma}\frac{\alpha}{\pi v}A_{\mu\nu}Z^{\mu\nu}h-\left(\kappa_u\sum_u\frac{m_f}{v}\bar{f}f + \kappa_d \sum_d \frac{m_f}{v}\bar{f}f + \kappa_\ell\sum_\ell \frac{m_f}{v}\bar{f}f\right)h
\end{align}
where, at tree level, we have
\begin{align}
\kappa_Z = \kappa_W = \sin(\beta-\alpha),\quad \kappa_u=\frac{\cos\alpha}{\sin\beta},\quad \kappa_{d,\ell}=-\frac{\sin\alpha}{\cos\beta}.
\end{align}
Note that $\kappa_g$, $\kappa_\gamma$ and $\kappa_{Z\gamma}$ are generated at one-loop. From above $\kappa$'s, the alignment limit can be identified as $\cos(\beta-\alpha)=0$\footnote{Another possibility $\sin(\beta-\alpha)=0$ is not considered here.}.

It is also important to have a discussion for the triple couplings among scalars. At the alignment limit, we have
\begin{align}
\lambda_{h\Phi\Phi}=-\frac{C_\Phi}{2v}(m_h^2+2m_\Phi^2 - \frac{2m_{12}^2}{\sin\beta\cos\beta}).
\end{align}
with $C_\Phi=2(1)$ for $\Phi=H^\pm(H,A)$. With degenerate masses $m_\Phi\equiv m_H=m_A=m_{H^\pm}$, we introduce a new parameter $\lambda$ as
\begin{align}
\lambda v^2\equiv m_\Phi^2 - \frac{m_{12}^2}{\sin\beta\cos\beta},
\end{align}
which is the parameter that enters the Higgs self-couplings and relevant for the loop corrections to the SM-like Higgs boson couplings. This parameter could be used interchangeably with $m_{12}^2$. For the rest of our analysis, $v=246$ GeV and $m_h=125$ GeV are used. The remaining free parameters are
\begin{align}
\tan\beta, \cos(\beta-\alpha), m_H, m_A, m_{H^\pm}, \lambda
\end{align}

\subsection{Loop corrections to Higgs couplings}
We define the normalized SM-like Higgs boson couplings including loop effects as
\begin{align}
\kappa_{\text{loop}}^{\text{2HDM}} \equiv \frac{g_{\rm  tree}^{\rm 2HDM}+g_{\rm loop}^{\rm 2HDM}}{g_{\rm tree}^{\rm SM}+g_{\rm loop}^{\rm SM}}
\end{align}

In our calculations, we adopt the on-shell renormalization scheme~\cite{FeynArts-SM}. The conventions for the renormalization constants and the renormalization conditions are mostly following Refs.~\cite{FeynArts-SM,Kanemura:2004mg}. All related counter terms, renormalization constants and renormalization conditions are implemented according to the on-shell scheme and incorporated into model files of 
{\tt FeynArts} \cite{Hahn:2000kx}\footnote{Note that in this scheme, there will be gauge-dependence in the calculation of the counter term of $\beta$~\cite{Freitas:2002um}. For convenience, we will adopt this convention and the Feynman-'t-Hooft gauge is used throughout the calculations. For more sophisticated gauge-independent renormalization scheme to deal with $\alpha$ and $\beta$, see~\cite{Krause:2016oke,Denner:2016etu,Altenkamp:2017ldc,Kanemura:2017wtm}. Corresponding implementations have been uploaded to \href{https://github.com/ycwu1030/THDMNLO_FA}{https://github.com/ycwu1030/THDMNLO\_FA}.}.
One-loop corrections are generated using {\tt FeynArts} and {\tt FormCalc}~\cite{Hahn:2016ebn} including all possible one-loop diagrams. {\tt FeynCalc}~\cite{Shtabovenko:2016sxi,Mertig:1990an} is also used to simplify the analytical expressions.   {\tt LoopTool}~\cite{Hahn:1998yk} is used to evaluate the numerical value of all the loop-induced amplitude. The numerical results have been cross-checked with another numerical program {\tt H-COUP}~\cite{Kanemura:2017gbi} in some cases.

Further, the 2HDM contributions to the oblique parameters are given by~\cite{He:2001tp}:

\begin{eqnarray}
\label{eqs:2HDM_EWPD}
\Delta\,S&=&\frac{1}{\pi\, m_Z^2} \Big\{  \Big[{\cal B}_{22}( m_Z^2\,; m_H^2\,, m_A^2 ) - {\cal B}_{22}( m_Z^2\,; m_{H^\pm}^2\,, m_{H^\pm}^2)\Big] \nonumber \\
&&+ \Big[ {\cal B}_{22}( m_Z^2\,; m_h^2\,, m_A^2 )- {\cal B}_{22}( m_Z^2\,; m_H^2\,, m_A^2 ) + {\cal B}_{22}( m_Z^2\,; m_Z^2\,, m_H^2 ) - {\cal B}_{22}( m_Z^2\,; m_Z^2\,, m_h^2 ) \nonumber \\
&& - m_Z^2 {\cal B}_0 ( m_Z\,; m_Z\,, m_H^2 ) + m_Z^2 {\cal B}_0 ( m_Z\,; m_Z\,, m_h^2 )    \Big] \cos^2(\beta-\alpha)  \Big\}\,,\\
\Delta T&=& \frac{1}{ 16\pi\, m_W^2\, s_W^2 } \Big\{   \Big[ F( m_{H^\pm}^2 \,, m_A^2 ) +  F( m_{H^\pm}^2\,, m_H^2) - F(m_A^2\,, m_H^2)  \Big]\nonumber \\
&&+   \Big[  F( m_{H^\pm}^2 \,, m_h^2 ) -  F( m_{H^\pm}^2\,, m_H^2) - F( m_A^2\,, m_h^2 )+ F(m_A^2\,, m_H^2)   \nonumber \\
&& + F( m_W^2\,, m_H^2) - F(m_W^2\,, m_h^2 ) - F(m_Z^2 \,, m_H^2 ) + F(m_Z^2\,, m_h^2) \nonumber \\
&& + 4 m_Z^2 \overline B_0 (m_Z^2\,, m_H^2\,, m_h^2)   - 4 m_W^2 \overline B_0 ( m_W^2\,, m_H^2\,, m_h^2 )  \Big] \cos^2(\beta-\alpha) \Big\}\,, \\
\Delta\,U&=& -\Delta\,S + \frac{1}{\pi m_W^2}\Big\{ \Big[{\cal B}_{22}(m_W^2,m_A^2,m_{H^\pm}^2)-2{\cal B}_{22}(m_W^2,m_{H^\pm}^2,m_{H^\pm}^2) + {\cal B}_{22}(m_W^2,m_H^2,m_{H^\pm}^2)\Big] \nonumber \\
&& + \Big[{\cal B}_{22}(m_W^2,m_h^2,m_{H^\pm}^2) - {\cal B}_{22}(m_W^2,m_H^2,m_{H^\pm}^2) + {\cal B}_{22}(m_W^2,m_W^2,m_H^2) - {\cal B}_{22}(m_W^2,m_W^2,m_h^2) \nonumber \\
&& - m_W^2{\cal B}_0(m_W^2,m_W^2,m_H^2) + m_W^2{\cal B}_0(m_W^2,m_W^2,m_h^2)\Big]\cos^2(\beta-\alpha) \Big\}\,,
\end{eqnarray}

\section{Global Fitting}
In our analysis, global fit is performed to get the constraints on the model parameters by constructing the $\chi^2$ with the profile likelihood method
\begin{align}
\chi^2 = \sum_i \frac{(\mu_i^{\text{BSM}}-\mu_i^{\text{obs}})^2}{\sigma_{\mu_i}^2},
\end{align} 
where $\mu_i^{\text{BSM}}=\frac{(\sigma\times{\rm BR})_{\text{BSM}}}{(\sigma\times{\rm BR})_{\text{SM}}}$, and $i$ runs over all available channels, the corresponding errors $\sigma_{\mu_i}$ for all these channels are listed in Tab.~\ref{tab:mu_precision}.

We also incorporate the Z-pole precision measurements by fitting into the oblique parameters $S$, $T$ and $U$ of which the $\chi^2$ reads
\begin{align}
\chi^2 = \sum_{ij}(X_i-\hat{X}_i)(V^{-1})_{ij}(X_j-\hat{X}_j),
\end{align}
with $X_i = (\Delta S, \Delta T, \Delta U)_{\text{2HDM}}$ being the predicted values in 2HDM, while $\hat{X}_i = (\Delta S, \Delta T, \Delta U)$ being the best-fit central value (assuming 0 for future measurements). $V$ is the covariance matrix with $V_{ij}=\sigma_i\rho_{ij}\sigma_j$. The corresponding errors $\sigma_i$ and the correlation matrix $\rho_{ij}$ are listed in Tab.~\ref{tab:STU} for both current measurement and ILC prospects.

\begin{figure}[!hbt]
\centering
\includegraphics[width=0.45\textwidth]{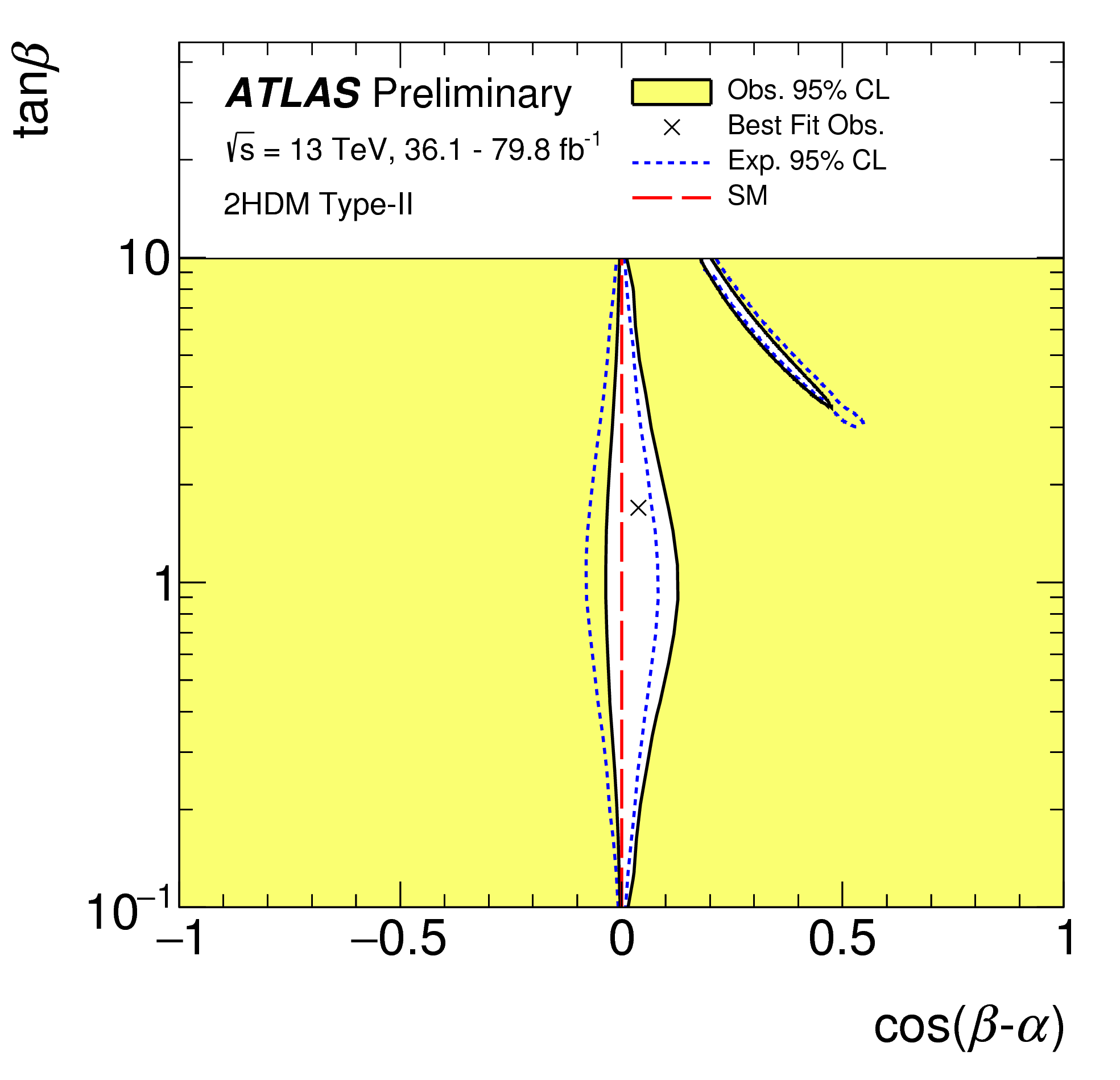}
\includegraphics[width=0.45\textwidth]{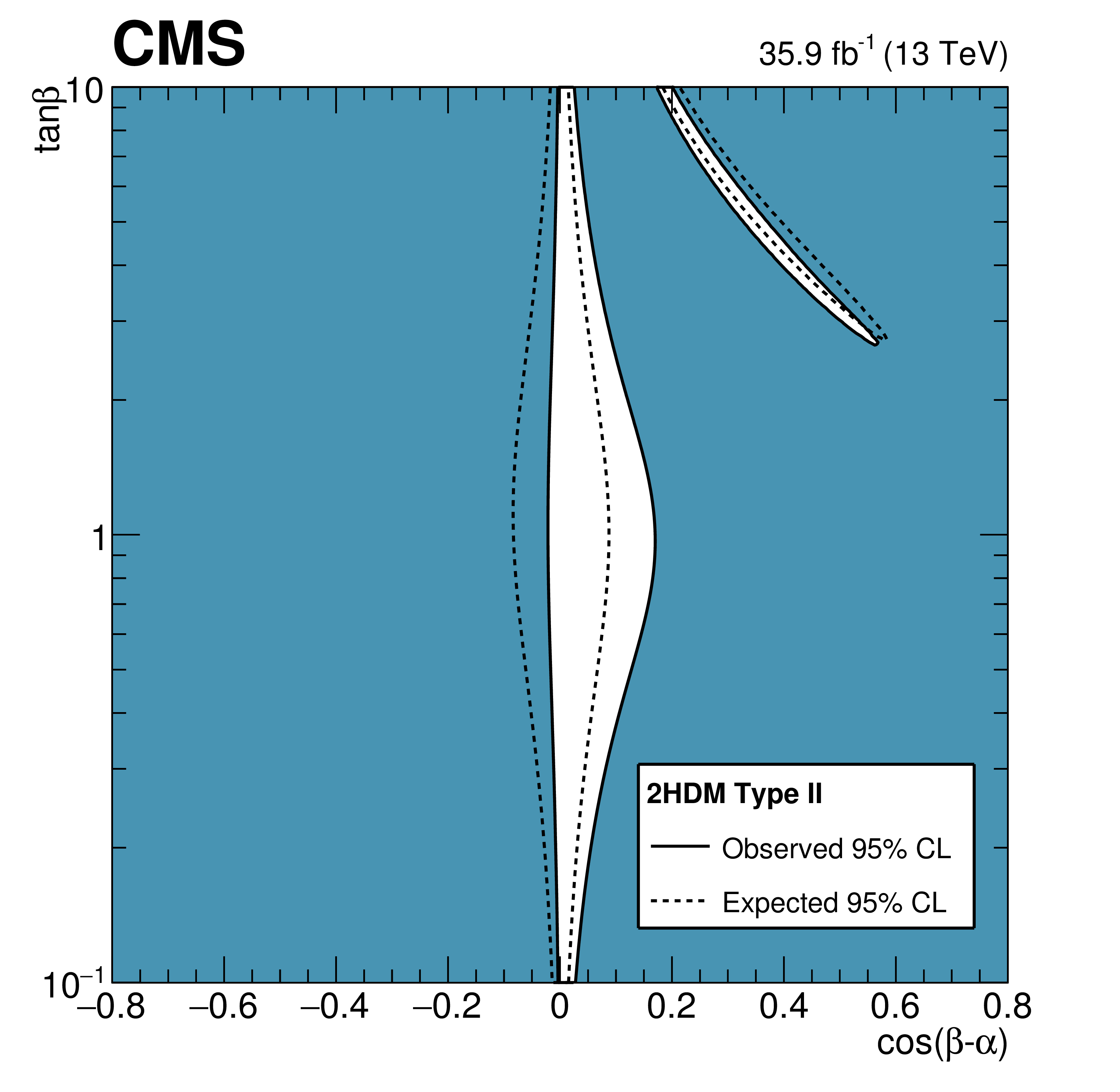}
\caption{Current constraints from ATLAS (left) and CMS (right) for the Type-II 2HDM in $\tan\beta$ vs. $\cos(\beta-\alpha)$ plane. Figures are getting from~\cite{ATLAS-CONF-2018-031,Sirunyan:2018koj}.}
\label{fig:CurrentTree}
\end{figure}

\begin{figure}[!hbt]
\centering
\includegraphics[width=0.6\textwidth]{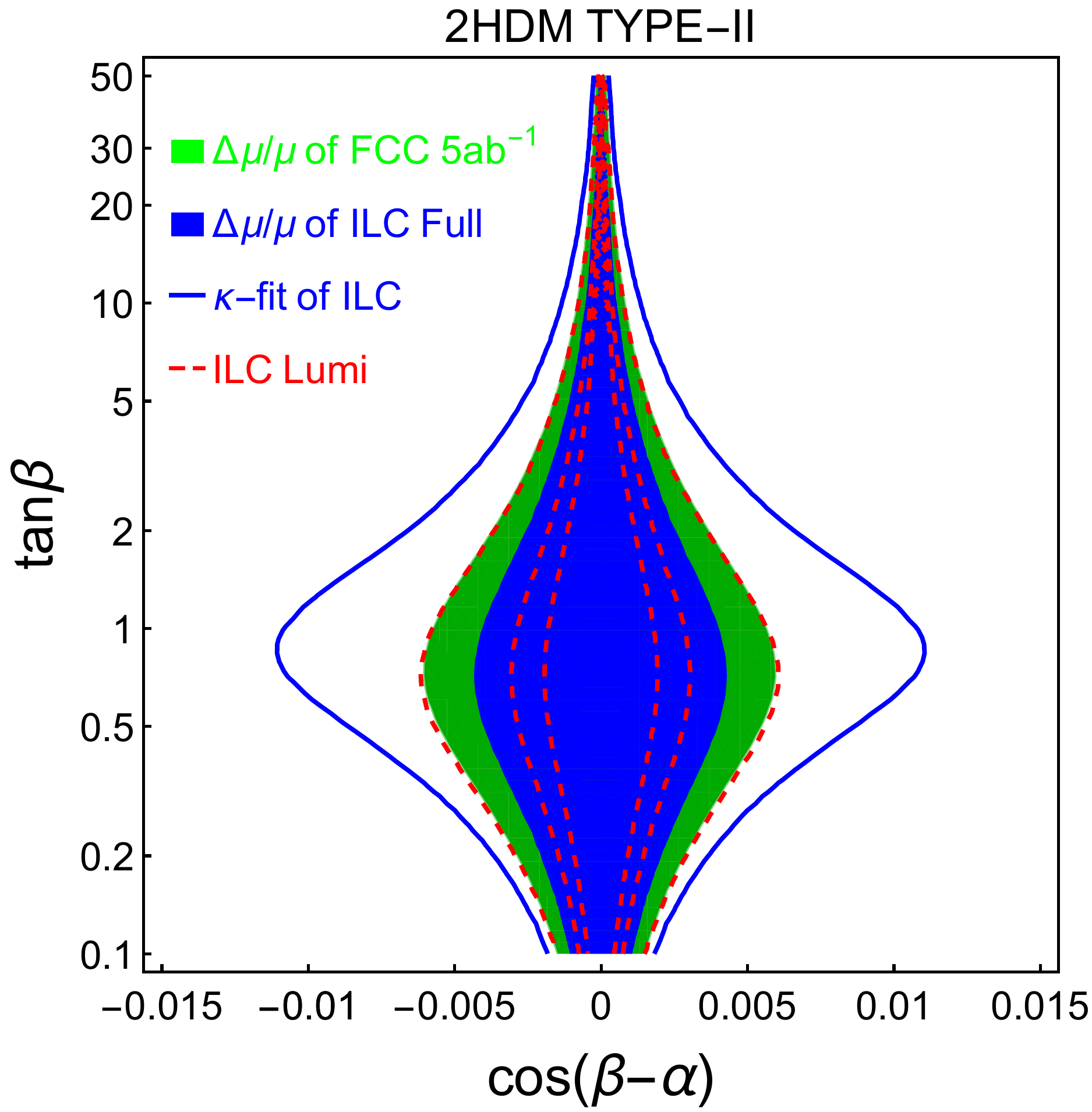}
\caption{The prospected constraints from ILC measurement for the Type-II 2HDM in $\tan\beta$ vs. $\cos(\beta-\alpha)$ plane.}
\label{fig:ILCTree}
\end{figure}

At tree level, the Higgs couplings only involve $\tan\beta$ and $\cos(\beta-\alpha)$. Hence, from Higgs measurement, we can only constraint the parameter space in $\tan\beta$ vs. $\cos(\beta-\alpha)$ plane. The results from ATLAS~\cite{ATLAS-CONF-2018-031} and CMS~\cite{Sirunyan:2018koj} are shown in Fig.~\ref{fig:CurrentTree}. Beside the extra arm corresponding to the wrong sign scenario, the current measurement strongly constrain the range of $\cos(\beta-\alpha)$. This can be further improved by the Higgs measurement at the ILC which can be seen in Fig.~\ref{fig:ILCTree}. From the fitting results, at tree level, we can already constrain the $\cos(\beta-\alpha)$ to be at least less than 0.01. 

Using $hZZ$ coupling as example, at tree level, the couplings is 
\begin{align}
\mathcal{L}_{hZZ} = \frac{m_Z^2}{v}s_{\beta-\alpha}hZ_\mu Z^\mu \approx \left(1-\frac{1}{2}c_{\beta-\alpha}^2\right)\frac{m_Z^2}{v}h Z_\mu Z^\mu.
\end{align}
Hence, the deviation from the SM value is $\frac{1}{2}c_{\beta-\alpha}^2\frac{m_Z^2}{v}$. With the constraints from the ILC prospect, the coupling deviation is at about $\mathcal{O}(10^{-4})$ level. In this case, the loop induced correction which is not vanished even in the alignment limit can not be ignored any more. Further, including loop correction which has heavy scalars running in the loop also provides the possibility to constrain the mass of heavy scalars as well as the triple scalar couplings. With this consideration, we will consider three cases in sequence:
\begin{itemize}
\item Tree-level alignment + Mass degenerate:
\begin{itemize}
\item $\cos(\beta-\alpha)=0$, $m_\Phi\equiv m_H=m_A = m_{H^\pm}$.
\end{itemize}
\item Non-alignment + Mass degenerate:
\begin{itemize}
\item $\cos(\beta-\alpha)\neq0$, $m_\Phi\equiv m_H=m_A = m_{H^\pm}$.
\end{itemize}
\item Non-alignment + Non-degenerate:
\begin{itemize}
\item $\cos(\beta-\alpha)\neq0$, $\Delta m_A = m_A-m_H$, $\Delta m_C = m_{H^\pm} - m_H$.
\end{itemize}
\end{itemize}

\subsection{Alignment and Mass Degenerate Case}
\begin{figure}[!hbt]
\centering
\includegraphics[width=0.6\textwidth]{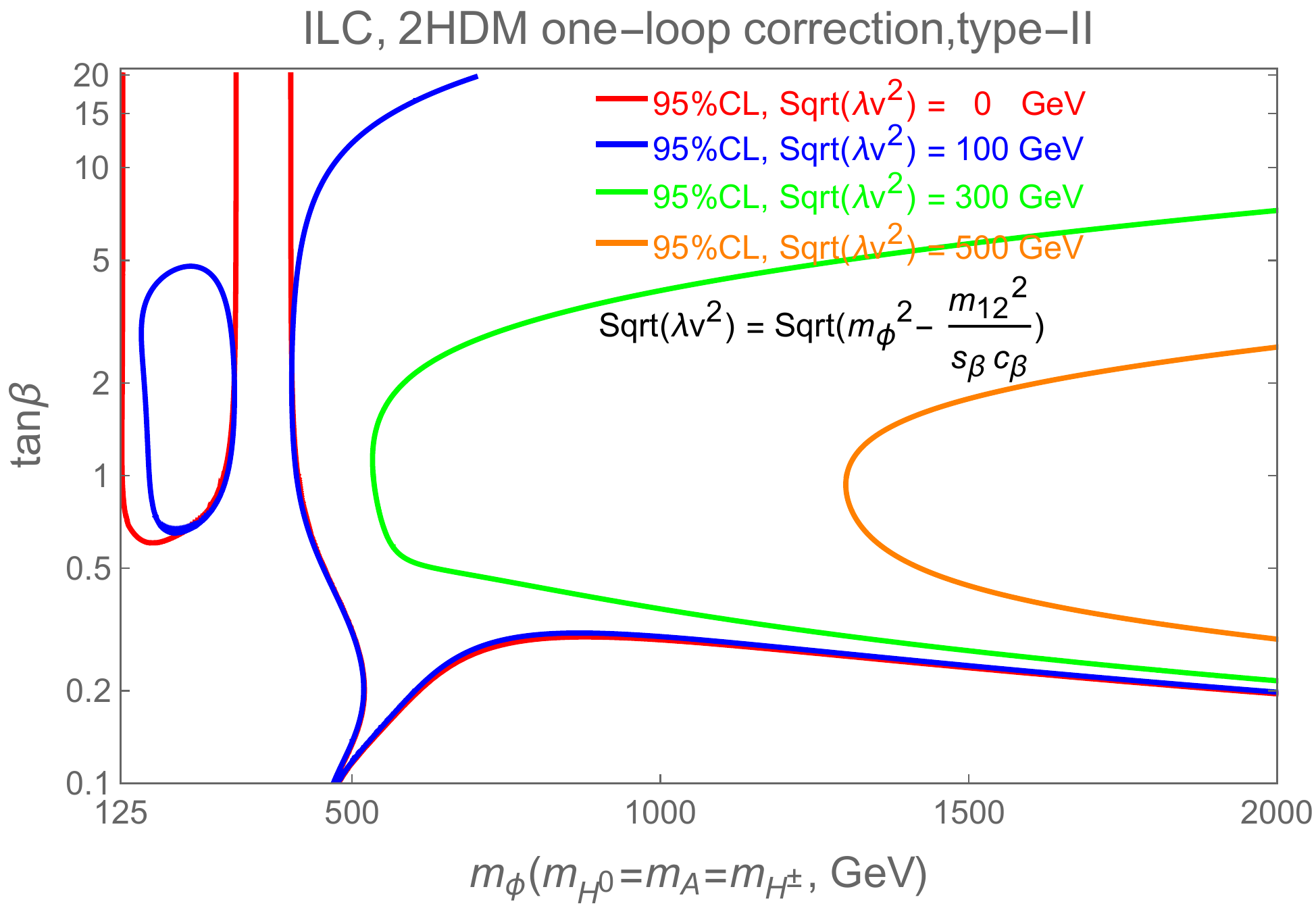}
\caption{The exclusion line from Higgs measurement in the $m_\Phi$-$\tan\beta$ plane for different choice of $\lambda$}
\label{fig:ILC_AliDege}
\end{figure}

The first case we consider is when $\cos(\beta-\alpha)=0$ and all the heavy scalars are degenerate in mass. In this case, the free parameters are $\tan\beta$, $m_\Phi$ and $\lambda$. The corresponding exclusion results are shown in Fig.~\ref{fig:ILC_AliDege} where lines with different colors represent the exclusions for different choice of $\lambda$ in $m_\Phi$-$\tan\beta$ plane. The region to the right of the lines and the upper left region are allowed. From this figure, we can see that when the triple scalar coupling is not strong, we still have large allowed region as shown by the red and blue curves. The gaps around 350 GeV come from the top-pair threshold. When the triple scalar coupling is strong, the exclusion curves provide lower limits on the heavy scalar mass which is consistent with our prospects. For $\sqrt{\lambda v^2} = 300$ GeV, a lower limit of 500 GeV on the mass is expected, while for $\sqrt{\lambda v^2} = 500$ GeV, the heavy scalar mass should be larger than 1200 GeV.

\subsection{Non-alignment and Mass Degenerate Case}
\begin{figure}[!hbt]
\centering
\includegraphics[width=0.45\textwidth]{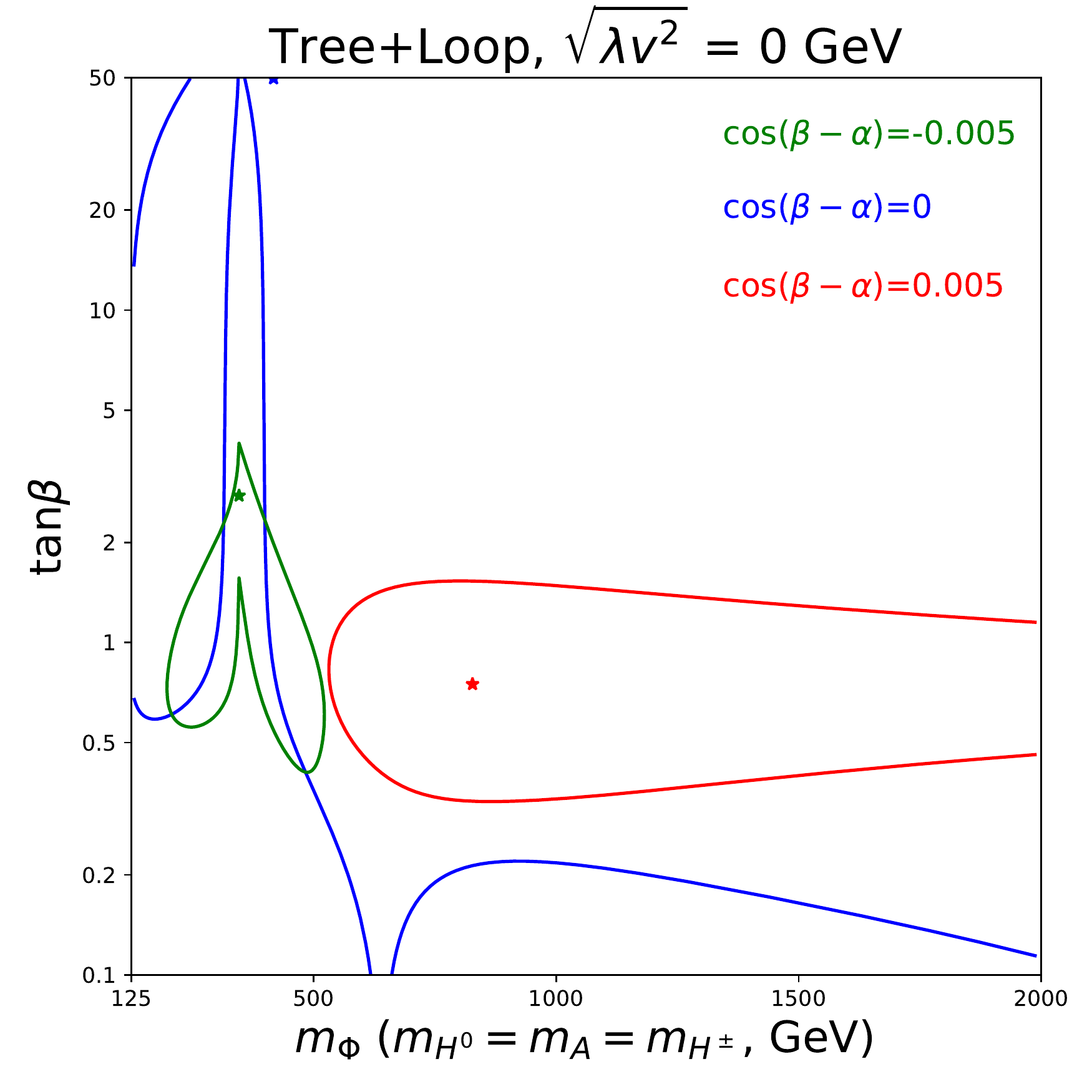}
\includegraphics[width=0.45\textwidth]{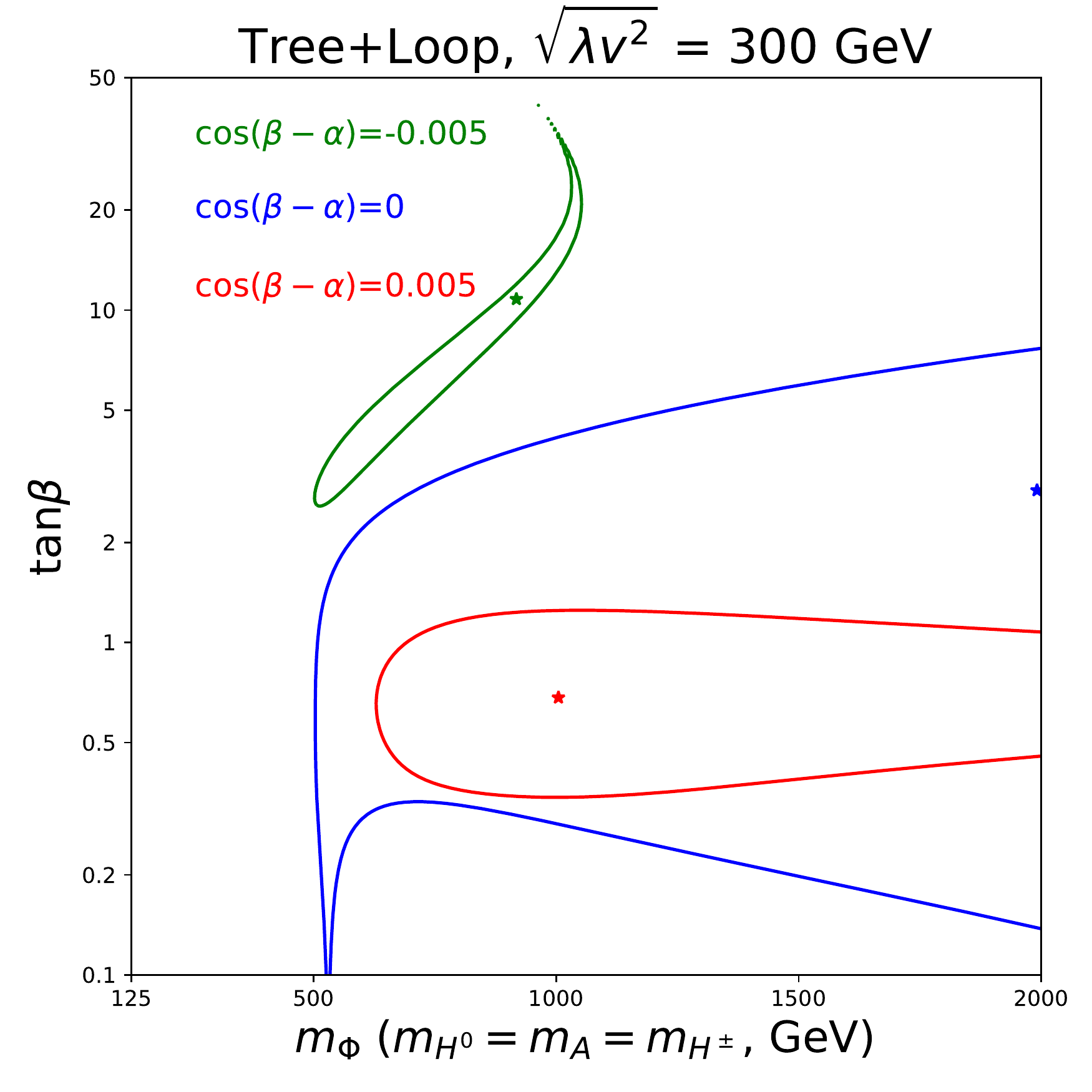}
\caption{The exclusion lines for different choice of $\cos(\beta-\alpha)$ in $m_\Phi$-$\tan\beta$ plane, the left panel is for $\sqrt{\lambda v^2} = 0$ GeV while the right panel is for $\sqrt{\lambda v^2} = 300$ GeV.}
\label{fig:ILC_TreeLoopDege_tanbvsmass}
\end{figure}

The second case we consider is when we deviate from the alignment limit but still assume that all heavy scalars are degenerated. In this case, the free parameters are $\tan\beta$, $\cos(\beta-\alpha)$, $m_\Phi$ and $\lambda$. We first show the result in Fig.~\ref{fig:ILC_TreeLoopDege_tanbvsmass} in the $m_\Phi$-$\tan\beta$ plane for different choice of $\cos(\beta-\alpha)$ (represented by lines with different colors) and of $\sqrt{\lambda v^2}$ (left panel: $\sqrt{\lambda v^2} = 0$ GeV and right panel: $\sqrt{\lambda v^2} = 300$ GeV). The constraints on other parameters are similar to the previous case, while it also shows different exclusion patterns for negative and positive choice of $\cos(\beta-\alpha)$. Note that, for the tree-level only results, Fig.~\ref{fig:ILCTree}, the constraints is symmetric for $\cos(\beta-\alpha)$. This shows the importance of the interference between loop corrections and the tree level couplings.

\begin{figure}[!hbt]
\centering
\includegraphics[width=0.54\textwidth]{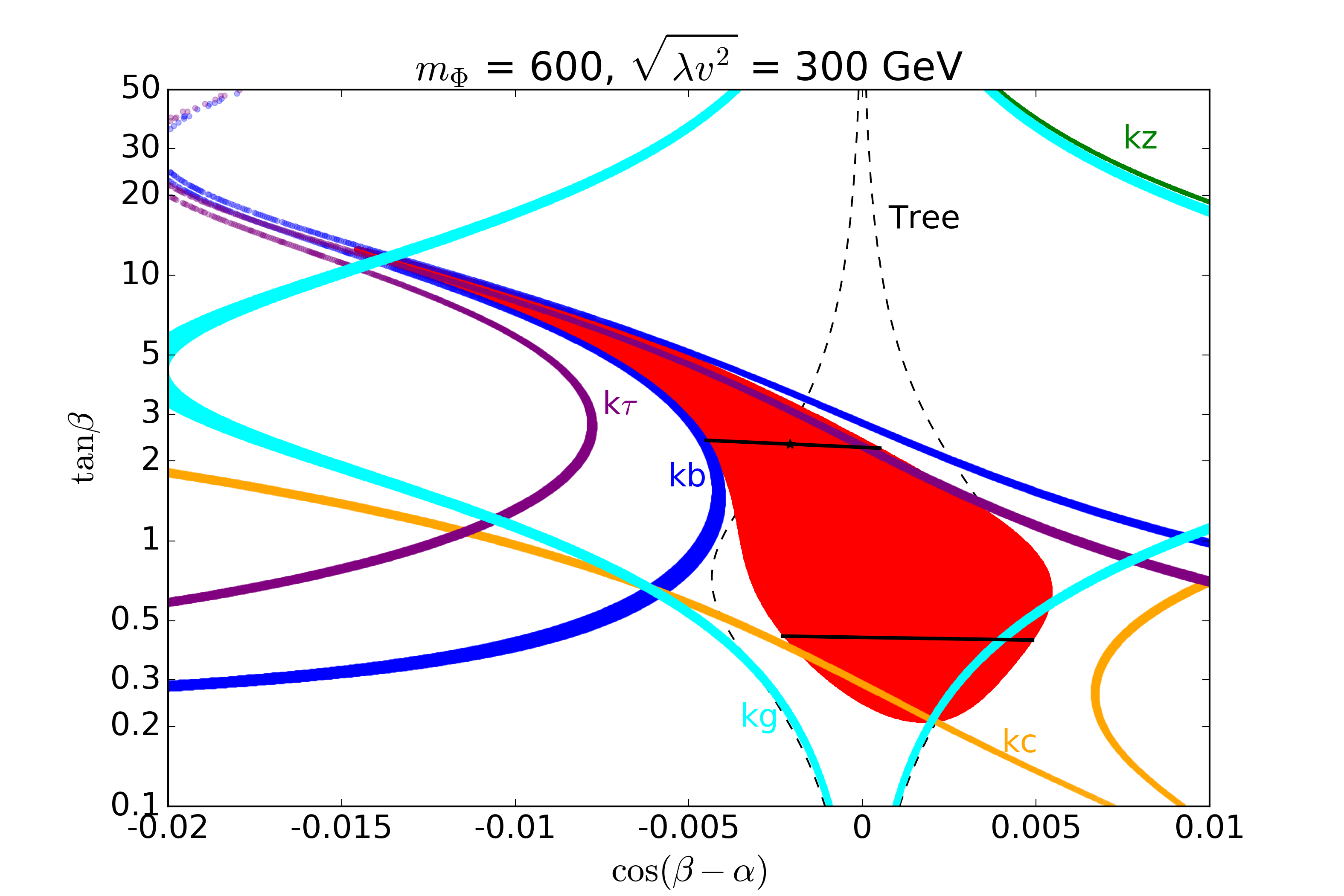}
\includegraphics[width=0.36\textwidth]{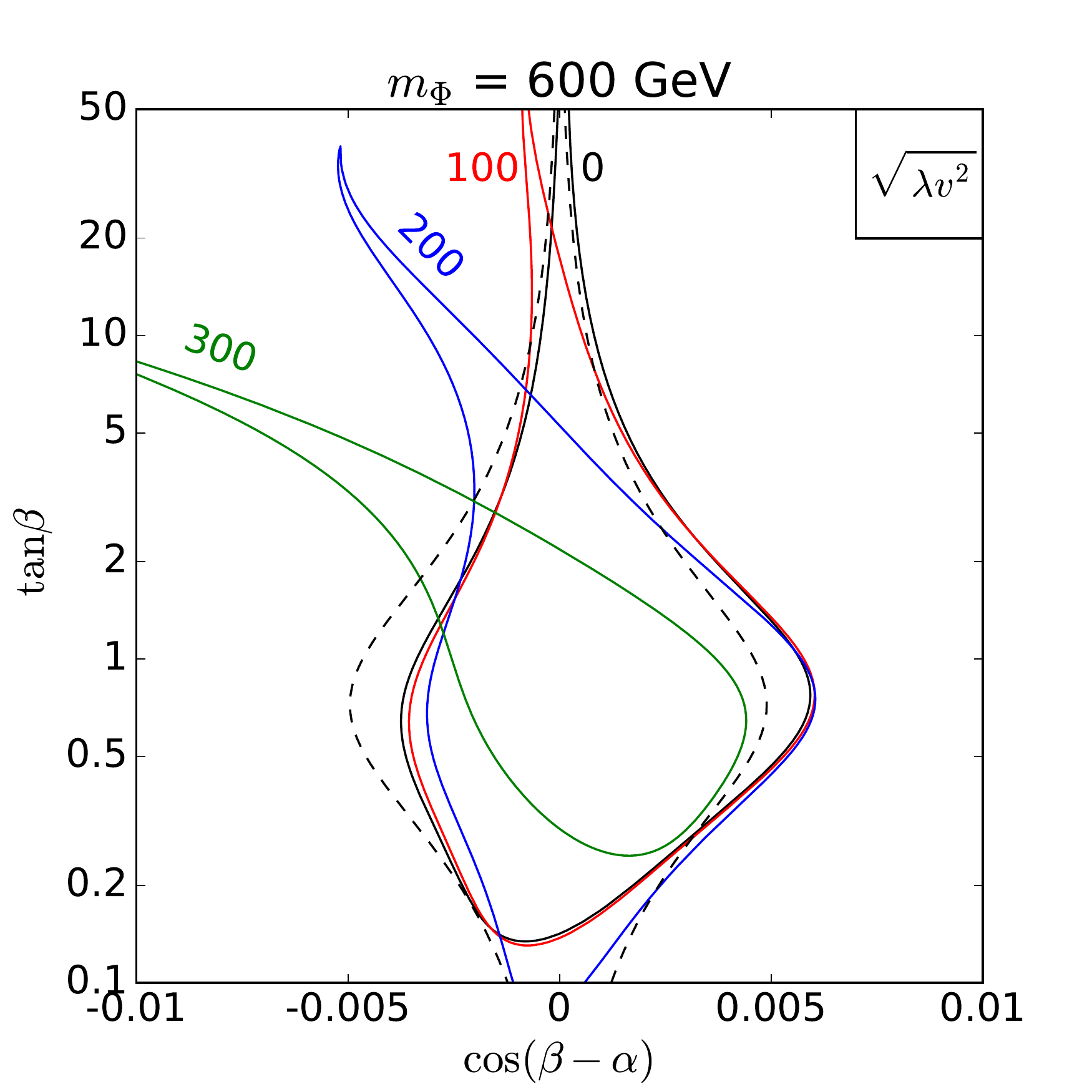}
\caption{The exclusion lines in the $\cos(\beta-\alpha)$-$\tan\beta$ plane for $m_\Phi = 600$ GeV. In the left panel, $\sqrt{\lambda v^2} = 300$ GeV is fixed. The red region is the global fitting allowed region, while lines with different colors are the constraints from different couplings as indicated by the label. The black dashed line represents the tree level result. In the right panel, only the global fitting allowed regions are shown for different choice of $\sqrt{\lambda v^2}$.}
\label{fig:ILC_TreeLoopDeg_tanbvscba}
\end{figure}

To make this more clearer, we show the results in the $\cos(\beta-\alpha)$-$\tan\beta$ plane in Fig.~\ref{fig:ILC_TreeLoopDeg_tanbvscba}. In the left panel, $\sqrt{\lambda v^2} = 300$ GeV is fixed. The red region is the global fitting allowed region, while lines with different colors are the constraints from different couplings as indicated by the label. We can see that at large $\tan\beta$ region, the constraints from $\kappa_\tau$ and $\kappa_b$ push the allowed region away from the tree level results. At low $\tan\beta$ region, the allowed region is cut off by $\kappa_c$ and $\kappa_g$. Note that we are still in the degenerate cases, with this small $\cos(\beta-\alpha)$ we are probing, the constraint from $\kappa_z$ is rather weak as shown by the green line in the upper right corner. The effects of $\sqrt{\lambda v^2}$ is shown in the right panel of Fig.~\ref{fig:ILC_TreeLoopDeg_tanbvscba}. We can see that the region will deviate from the tree level cases with increase of $\sqrt{\lambda v^2}$. Note that there is no more allowed region for larger value of $\sqrt{\lambda v^2}$ which put an upper limit on $\lambda$.

\subsection{Non-alignment and Non-degenerate Case}
\begin{figure}[!hbt]
\centering
\includegraphics[width=0.6\textwidth, trim=120 120 120 120]{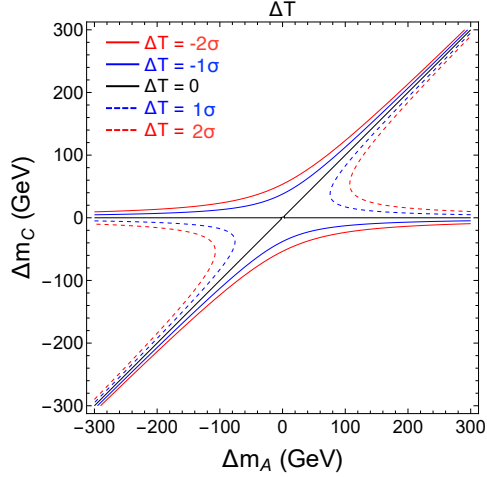}
\caption{The constraints from T-parameter in $m_A$-$m_C$ plane. The blue and red lines represent the 1-$\sigma$  and 2-$\sigma$ region respectively.}
\label{fig:TPara}
\end{figure}

The last case we consider is the most general one. We allow both the deviation from the alignment and also the mass splittings. The new free parameters are the mass splittings among different scalars: $\Delta m_A = m_A - m_H$, $\Delta m_C = m_{H^\pm} - m_H$. The constraints on other parameters are similar to previous case. Hence, we focus on the constraints on mass differences here. When allowing mass splitting, the Z-pole measurements will also provide stringent constraints. The constraints from the T parameter is shown in Fig.~\ref{fig:TPara} in $m_A$-$m_C$ plane. From this plot we can see that the Z-pole measurement has already put a strong constraint on the mass difference. Hence, a natural question would be whether the Higgs measurement can provide further information on the mass difference.

\begin{figure}[!hbt]
\centering
\includegraphics[width=0.9\textwidth]{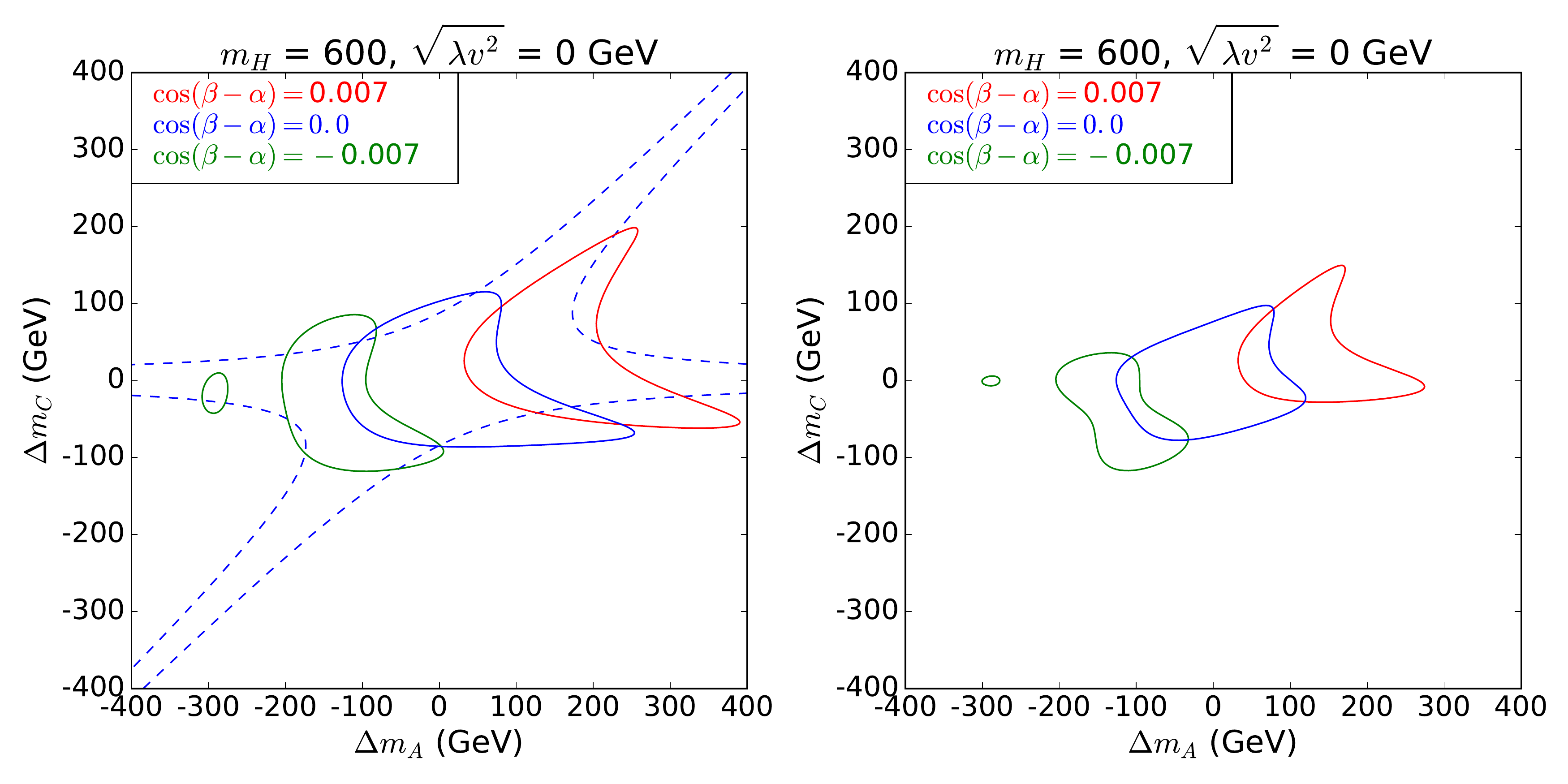}\\
\includegraphics[width=0.9\textwidth]{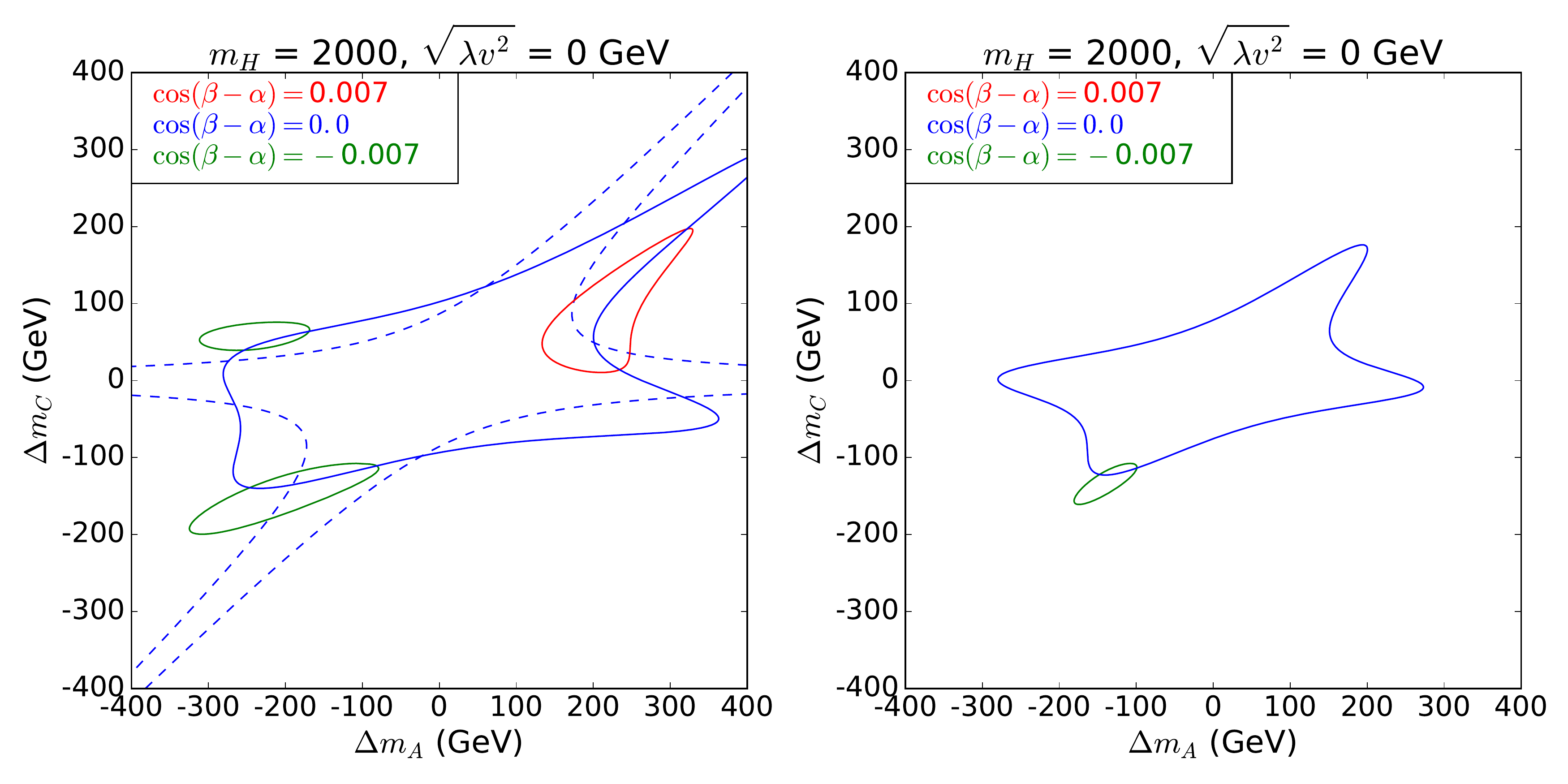}
\caption{The constraints from Higgs measurement (left column) and Higgs measurement combined with Z-pole measurement (right column) in $\Delta m_A$-$\Delta m_C$ plane. The mass are set by $m_H = 600$ and 2000 GeV for upper two and lower two panels respectively. Lines with different colors are for different choices of $\cos(\beta-\alpha)$ as indicated in the legend. The dashed contours in the left column are the Z-pole constraints which barely changes with $\cos(\beta-\alpha)$ in this range.}
\label{fig:DeltaMass}
\end{figure}

The results from the Higgs measurement are shown in Fig.~\ref{fig:DeltaMass}. In the left column, the individual constraints from Higgs measurement (solid lines) and Z-pole measurement (dashed line) are shown. We can see that both Higgs measurement and Z-pole measurement can constrain the mass splitting, while these two constraints are mis-aligned. Furthermore, the Higgs measurement is quite sensitive to the value of $\cos(\beta-\alpha)$. However, the T parameter constraints here merely changes for the range of the $\cos(\beta-\alpha)$ we considered here. When combining these, the allowed region is further shrunk as shown in the right column of Fig.~\ref{fig:DeltaMass}. This result clearly shows the complementarity between the Higgs measurement and Z-pole measurement. For $\cos(\beta-\alpha)=0$ (blue lines), increasing the mass of the heavy scalar leads to larger allowed region which is consistent with the decoupling limit. However, when we deviate from $\cos(\beta-\alpha)=0$ ($\pm 0.007$ as shown by red and green lines in the plots), the allowed the region is even smaller for larger mass. This is due to the fact that when we approach decoupling limit, the parameter space will also be pushed into alignment limits, otherwise we suffer from either the unitarity/perturbativity problem or large correction induced by triple scalar couplings.

\section{Summary}

In this paper, we examined the impacts of the precision measurements of the SM parameters at the proposed Higgs factories using Type-II 2HDM as an example. At tree level, the current measurement has already push the model into the alignment limit region. The proposed Higgs factories can do even better with clean background. In this region, for some coupling, the loop corrections can not be ignored any more. When including loop induced contributions, we can further set bounds on the heavy scalar masses as well as the mass splitting. The triple scalar couping is constrained as well. In this sense, the Higgs precision measurement is complementary to the Z-pole measurement. The combination of these two kinds measurements can certainly provide more information about the new physics. 

\begin{acknowledgments}

%%%%%%%%%%%%%%%%%%%%%%%%%%%%%%%%%%%%%%%%%%%%%%%%%%%%%%%%%%%%%%%%%%%%%%%%%%%%%%%%%%%%%%

We would like to thank Han Yuan and Huanian Zhang for collaboration at the early stage of this project.  We would also like to thank Liantao Wang and Manqi Ruan for valuable discussions.
NC is supported by the National Natural Science Foundation of China (under Grant No. 11575176) and Center for Future High Energy Physics (CFHEP).
TH is supported in part by the U.S.~Department of Energy under
grant No.~DE-FG02-95ER40896 and by the PITT PACC. SS is supported  by the Department of Energy under Grant No.~DE-FG02-13ER41976/DE-SC0009913.
WS were supported in part by the National Natural Science Foundation of China (NNSFC) under grant No.~11675242. YW is supported by the Natural Sciences and Engineering Research Council of Canada (NSERC). TH also acknowledges the hospitality of the Aspen Center for Physics, which is supported by National Science Foundation grant PHY-1607611.

 \end{acknowledgments}

\bibliographystyle{bibsty}
\bibliography{references}

\end{document}